# A FRAMEWORK FOR SIMULATING SYSTEMIC RISK AND ITS APPLICATION TO THE SOUTH AFRICAN BANKING SECTOR

**By NM Walters, FJC Beyers, AJ van Zyl & RJ van den Heever**

ABSTRACT

We present a network-based framework for simulating systemic risk that considers shock propagation in banking systems. In particular, the framework allows the modeller to reflect a top-down framework where a shock to one bank in the system affects the solvency and liquidity position of other banks, through systemic market risks and consequential liquidity strains. We illustrate the framework with an application using South African bank balance sheet data. Spikes in simulated assessments of systemic risk agree closely with spikes in documented subjective assessments of this risk. This indicates that network models can be useful for monitoring systemic risk levels. The model results are sensitive to liquidity risk and market sentiment and therefore the related parameters are important considerations when using a network approach to systemic risk modelling.

CONTACT DETAILS

## 1. INTRODUCTION

1.1    Systemic risk and the spread of financial contagion are important considerations for regulators tasked with overseeing stability of banking systems. Banking systems are at the core of a well-functioning financial system. A breakdown of the system would hinder economic growth, which in turn may cause permanent damage to the economy (Cerra & Saxena, 2017). Therefore, it is important for regulators to prevent such a breakdown from being triggered. Regulatory intervention at a late stage could prove to be costlier than intervention at an earlier stage. The burden of costly bailouts by the regulator are ultimately borne by the taxpayers, which negatively affects the economy. On the other hand, if banks are allowed to fail without any intervention, the economy can be strained by losses on investors' deposits, rising interest rates, possible bank runs etc. Monitoring the level of systemic risk in a financial system is therefore crucial for ensuring long-term stability and growth of an economy.

1.2    Liquidity and market sentiment are two key requirements for a working banking system that are also closely related. During times of economic distress, a lack of trust translates into a reluctance of non-bank financial institutions to renew funding to banks. They then impose more stringent lending requirements, which leads to increased risk premia on loans and debentures thereby increasing banks' wholesale funding costs. The higher interest rates charged on servicing new debt means that additional assets may need to be liquidated to service the debt or a reduction in asset origination, reducing (shrinking) the balance sheet sizes of the affected banks. This puts a strain on those banks' liquidity positions as the maturity mismatch between short-term liabilities and assets increases. Ultimately, when the funding costs become unsustainably high the bank may be forced to call in loans or liquidate assets prematurely. This, together with the increased funding costs can substantially reduce the bank's profitability and hence its retained earnings. This in turn reduces its Tier I capital, which may lead to solvency problems (Furceri & Mourougane, 2009). This creates a spiral of distrust.

1.3     The complex nature of banking systems remains difficult to replicate and model precisely. Bottom-up approaches using integrated modelling frameworks are very useful, yet they are difficult to calibrate, expensive and not readily available. This is because in practice, such an approach would involve the regulator providing a specified scenario to all banks, after which the banks quantify their own risk position so that the regulator can then aggregate the risk positions (Borio, Drehmann & Tsatsaronis, 2014). It is therefore of interest to find simplified models that consider the entire system from the start and can detect changes in systemic risk. We contribute to this by showing that network models of systemic risk can satisfy this requirement to a large extent. We illustrate how such a top-down model can be used, by applying it to real-world balance sheet data and showing that changes in risk are detected under times of market stress for various network structures. We turn our attention to problems in rolling forward short-term debt that is caused by frictions such as a lack of trust in the system.

1.4     The chain of events that we aim to model is as follows: One bank in the system experiences solvency problems, which may arise because of a significant increase in impairments from non-performing loans. This could be because of a number of causes such as unsustainable lending practices or a disruption in its target market (such as the mine closures experienced in South Africa). It is important to note that the applied model does not require us to specify the event that leads to the initial bank's default, nor do we attempt to model it. The equity of the aforementioned bank then declines, and shareholders need to absorb the losses (followed by other subordinated creditors). Now there are two key potential effects on the banking system. Firstly, other banks' balance sheets may be affected through a revaluation of assets and impairment provisions and they may need to raise additional impairment provisions (e.g. if the initial bank's troubles were due to increased impairments on a specific type of loan book, other banks may need to raise their impairment provisions for similar books to account for an anticipated rise in impairments). Another possibility is that the bank may ultimately need to resort to forced sales to generate liquidity. The increased supply of those assets in the market may depress their market value, leading to some (albeit limited) mark-to-market losses for other banks holding similar assets. For this study, the distinction between these possibilities (and hence the effect of the initial default on the banking or trading books of other banks) is not explicitly made. Here, we assume a net reduction in the balance sheets of other banks takes place which could be due to an increase in required reserves, or to a combination of this and mark-to-market losses. This approach is adopted to keep the model simple and consistent with existing models in the literature (see for example Nier et al. (2008), Gai & Kapadia (2010), May & Arinaminpathy (2010) Arinaminpathy, Kapadia & May (2012)). It is worthwhile to note that the narrative of these papers focusses on the fire-sale aspect of this contagion mechanism. However, as this is a practical application within a specific financial environment, we include losses due to raised provisions on the banking book in this mechanism as well. The second effect of the initially troubled bank affects the liability side of other banks' balance sheets and is more likely to lead to contagion. Funders' trust in the ability of banks to service their debt may decline as they become incapable of distinguishing between financially sound and troubled banks. This leads to liquidity issues, as the cost of rolling forward short-term debt increases for the affected banks as non-bank financial institutions are reluctant to renew their loans. Banks need to roll forward their short-term debt as they usually invest in long-term assets and take short-term deposits from funders. This gives them the needed liquidity at a low funding cost under normal circumstances. Banks may then be forced to sell assets below their market value to generate liquid funds and avoid maturity mismatches on their balance sheets.

1.5     As a top-down model, we propose the application of network theory. It has been applied in a wide variety of disciplines including sociology, computer science, epidemiology, biology, economics and finance. Network models of systemic risk have been developed in the literature over the past decade (see Upper & Worms (2004) and Chinazzi & Fagiolo (2013) for surveys). It involves representing the banking system as a network of interconnected agents, where interactions between banks are modelled explicitly. Allen & Babus (2008) explain why network theory is a useful tool for understanding and analysing systemic risk in financial systems and provides a meaningful way of analysing connections between them. They argue that network theory is instrumental in investigating financial stability by considering how a single institution can cause risk to the entire system. As shown by Georg (2011) and Ladley (2013), network theory can also be used to investigate the effect of common system-wide shocks.

## 2.     RELATED LITERATURE

### 2.1     BACKGROUND TO NETWORK MODELS

2.1.1.   A network is a system of $N$ interacting agents (called nodes), where the interactions between them form links between the nodes (called edges). It can be represented as graph, which is a pair $G = (V, E)$, where $V = \{1, 2, \dots, N\}$ denotes the collection of nodes and $E = \{\{i, j\}\}$ is the collection of edges. An edge $\{i, j\}$ is a collection of two nodes $i, j \in V$ that are connected via a link in the network. In this application, we use directed graphs. Here, the collection $\{i, j\}$ is ordered and each edge starts at the first node $i$ and ends at the second node $j$. Figures Figure *1* and 2 below illustrate the difference between undirected and directed graphs.

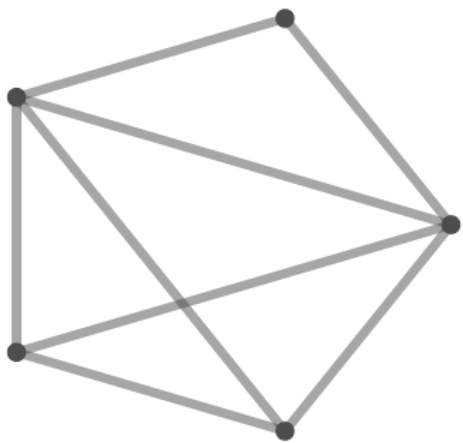
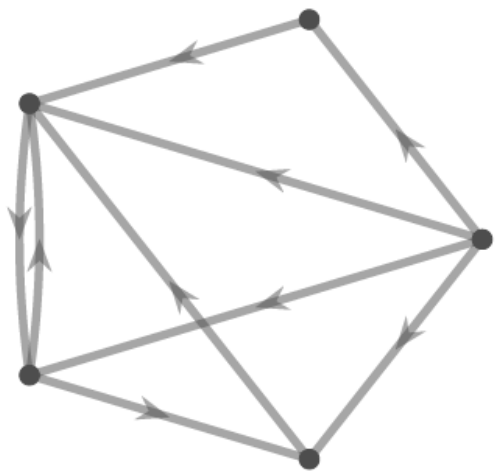

Figure 1. Illustration of an undirected graph          Figure 2. Illustration of a directed graph

2.1.2.   A commonly used method of modelling the edges between nodes is based on one of the earliest probability models of a graph, studied by Erdös & Rényi (1959). Here, each edge in the network is present with a fixed probability $p$, referred to as the Erdös-Rényi probability. The resulting graph is called an Erdös-Rényi network. It can be used for both directed and undirected graphs. We compare this model with extensions of it where the probability that a directed edge exists from node $i$ to $j$, $p(i,j)$, is dependent on the nodes $i$ and $j$. Note that for this application, $p(i,i) = 0$ for all $i \in V$, and it is said that the graph does not contain any loops (i.e. an edge cannot start and end at the same node).

2.1.3.   The banks are represented by nodes and the paths through which the problems of one bank spill over to others are represented by the edges. The edges need to be directed to take account of the direction in which the losses propagate through the system. The interpretation of the edges as mechanisms through which uncertainty is channelled is discussed further in section 3.2.

2.1.4.   The average probability that any node $i$ is connected to another node $j \neq i$ is given by $\bar{p} = \frac{1}{n(n-1)}\sum_{i \neq j} p(i,j)$. It can be used as a measure of the interconnectedness of a network. We also use it to standardise different extensions of the Erdös-Rényi network as described in Appendix C. The transmission of losses in the network as described in section 3.2 makes use of nodes' shortest paths. The shortest path from a node $i$ to another node $j$, say $d(i,j)$, is the smallest number of edges that can be used to travel from node $i$ to node $j$. It is used as a measure of distance between two nodes in a network and takes account of the edge directions.

2.1.5.   The way that banks are connected to one another via edges in a network is referred to as the structure of the network. Real-world systems are often more complex than purely random networks such as the standard Erdös-Rényi network (Kim & Wilhelm, 2008). They often exhibit characteristics such as low average shortest paths and nodes that are clustered together (called the small-world property). The Watts-Strogatz model (Watts & Strogatz, 1998) can be used to generate such a network. Other commonly used network models include those where nodes have a power law distribution of the number of edges connected to them (referred to as a power law degree distribution). The Barabásie-Albert model (Albert & Barabásie, 2002) can be used to generate networks with this property.

2.1.6.   Such power law distributions are of interest to large networks which that may exhibit this behaviour (Boss et al., 2004, Gabrieli, 2011, Santos & Cont, 2010, Cont, Moussa & Santos, 2012). For smaller networks such as the South African system, a power law distribution is theoretically unsuitable and difficult to test for. This paper addresses this shortcoming by investigating network structures that are applicable to smaller networks. We restrict ourselves to models where the probabilities $p(i,j)$ are functions of properties of the nodes $i$ and $j$. This is done firstly to enable us to test a wide variety of structures that are comparable to one another, since they are simulated in a similar way. Secondly, since the $p(i,j)$ probabilities are dependent on the nodes $i$ and $j$, it enables us to explicitly take account of differences between banks. Finally, models with power law degree distributions do not make much sense for networks with few nodes such the one we consider here.

## 2.2   RELEVANT LITERATURE

2.2.1.   Methods for measuring systemic risk can generally be categorised according to whether they are based on market information (i.e. asset prices) or balance sheet information. The studies by He & Krishnamurthy (2014) and Hautsch, Schaumburg & Schienle (2015) are examples of the former, whilst network models usually make use of balance sheet information. As methods based on market information rely on assumptions of market efficiency, this must be taken into account when interpreting the results, which is not the case for balance sheet driven models. Another advantage of using balance sheet data is that it is possible to separate the effects of systemic risk from any mitigating actions by regulators, whereas asset prices usually implicitly take account of the possibility of regulatory intervention (Birchler & Facchinetti, 2007, Alessandri et al., 2009). Network models have a further advantage of emphasizing the ways in which banks influence one another. Such models provide a clear distinction between individual entities and the financial network as a whole (Bisias et al., 2012).

2.2.2.   It is common for network models of systemic risk to have the edges between the nodes represent interbank assets and liabilities. Such models assume that whenever a bank in the system defaults, it cannot honour its commitments to its creditors, and hence defaults on its interbank commitments. We note that in the South African banking environment, the hierarchy of interbank loans compared to other unsecured debt is not well-defined. Therefore, in the event of a bank's default, other unsecured liabilities may be subjected to bail-in before interbank liabilities. That is why we take a different approach in this study, and do not model the interbank lending relationships.Instead, we consider contagion mechanisms applicable to any jurisdiction, namely how a loss of trust in the market may spill over to other banks, creating uncertainty and a difficulty to roll forward short-term debt.

2.2.3.   Regardless of what the edges represent, network models are flexible enough to be used in a range of different circumstances. For example, Markose, Giansante & Shaghaghi (2012) investigate the network of exposures generated by credit default swaps of the United States of America, Battiston et al. (2012) study the loans generated by the US Federal Reserve Bank emergency loans program, and Garratt, Mahadeva & Svirydzenka (2011) consider international banking exposures.

2.2.4.   When modelling systemic risk in financial systems, it is important to include other channels of contagion such as market liquidity risk (Upper, 2011). Chen, Liu & Yao (2016) find that market liquidity risk has a significant effect on systemic risk in a network setting. Network studies of systemic risk that incorporate market liquidity risk include those by Gai & Kapadia (2010), Roukny et al. (2013) and May & Arinaminpathy (2010), while Gai, Haldane & Kapadia (2011) focus on funding liquidity risk by incorporating haircuts to short term debt.

2.2.5.   We explicitly account for this by including factors representing market liquidity risk for short-, medium- and long-term assets. Short-term assets are defined to have a maturity of less than a month, and medium-term assets a maturity of more than a month and less than a year. Assets with a maturity of more than a year are deemed long-term. Network models generally assume simplified balance sheet structures to facilitate the modelling of key balance sheet components (see for example Nier et al. (2008), Gai & Kapadia (2010), Arinaminpathy, Kapadia & May (2012) and Cont, Moussa & Santos (2012)). A similar approach is taken by this study, with the simplified balance sheet illustrated in Table 1. The modelling of all channels of contagion is explained in section 3.2.

Table 1. Illustration of a simplified balance sheet

| Assets | Equity & liabilities |
| --- | --- |
| Short-term assets | Capital |
| Medium-term assets | Other equity & liabilities |
| Long-term assets | |

2.2.6.   One difficulty arising from network models is the specification of the links between banks, since it is not possible to determine the paths through which losses will spread. Empirical network studies that focus on interbank lending as the direct contagion channel can use maximum entropy estimation techniques to estimate connections between banks (Upper & Worms, 2004). However, this estimation method can lead to inaccuracies when assessing systemic risk (Mistrulli, 2011) and makes use of each bank's total interbank assets and loans. It is therefore not appropriate to use this for our contagion mechanism.

2.2.7.   The process of deciding which banks to connect to one another should consider the definition of the edges in the network. For this to make sense, the formation of edges should ideally be consistent with the event that initiated the contagion in the first place. For example, banks that are heavily exposed to the mining sector may experience a loss of investor sentiment as a result of sudden mine closures. From a modelling perspective, it is impractical to try to account for all possible bottom-up contagion events, seeing that this would need to cover a wide range of exposures and their associated risks. For this reason, we consider a range of different top-down network structures when modelling systemic risk. Furthermore, we note that different network structures may exhibit different levels of risk and the effect of changes to network characteristics is dependent on the chosen structure (Gai, Haldane & Kapadia, 2011, Georg, 2011, Krause & Giansante, 2012). As the true network structure is unknown, it is important to investigate how changes in the structure can affect the modelling of systemic risk.

2.2.8.   The structures investigated here are explained in section 3.3, some of which facilitate the modelling of core-peripheral networks. These are networks that consist of a small number of tightly connected 'core' banks and numerous sparsely connected 'peripheral' banks. Iori et al. (2008) and Fricke & Lux (2015) find evidence that real-world interbank networks exhibit this behaviour (Hüser, 2015, Glasserman & Young, 2016). Even though we do not model interbank lending relationships, it is possible that losses due to investor sentiment follow a similar pattern. Therefore, such structures are included for consideration in this study.

2.2.9.   We contribute to the growing body of empirical analyses of banking networks (see e.g. the work by Georg & Brink (2011), Huang, Zhou & Zhu (2012), Vallascas & Keasey (2012), Upper & Worms (2004), Martinez-Jaramillo et al. (2014) and Boss et al. (2004) to name but a few) by applying a network model of market sentiment to South African bank balance sheet data. The results of the model are used to assess whether the model can be used to monitor systemic risk levels by capturing increases in systemic risk during stressed market conditions. This is done by considering systemic risk at different points in time during which incidents occurred that adversely affected the local economy.

2.2.10. As the focus of this paper is on network mechanisms alone, it does not incorporate central bank activity or macro-economic factors. This is because their influence on the results would obscure effects relating to the network model itself and the differences in network structure. Central bank activity needs to be excluded as this research is done from the regulator's point of view and its reaction to bank failures are impossible to predict beforehand. The results would therefore be skewed by any assumptions made regarding central bank responses. From a practical perspective, regulators assessing policy responses to banking crises should compare the cost of intervention to the cost of not intervening, since both options can bear a high cost (Furceri & Mourougane, 2009). A useful area of future research would be to consider a set of policy responses and investigate how the network structure affects the risks borne by various stakeholders such as depositors, taxpayers and other banks. One can then consider the risk borne by each stakeholder when the risk is ignored and when the regulator intervenes. While it is possible and desirable to embed network models into larger macro-economic models (see for example Georg & Poschmann (2010) and Aikman et al. (2009)), it is specifically excluded here since one aspect of the study is to assess the performance of pure network models. Therefore, as this is a gross risk assessment, the inclusion of macro-economic effects may influence the conclusions unduly.

2.2.11. To summarise, this paper aims to make the following contributions:

(a) A top-down network approach is used to model systemic risk over time in the South African banking system. We investigate whether this model is capable of monitoring systemic risk by detecting instances of market turmoil.

(b) We introduce a novel contagion mechanism that focuses on market sentiment.

(c) The effect of the network structure on the results is investigated. Since the actual network structure is unknown, it is important to obtain insight into how sensitive the results are to the choice of network structure.

2.2.12. The remainder of this paper is structured as follows: Section 3 explains how the balance sheets are constructed, discusses the modelling procedure, and presents the different network structures. The results obtained by applying the model to the South African system are presented in section 4, after which section 5 concludes.

## 3. DATA AND NETWORK DESCRIPTION

### 3.1 DATA AND BALANCE SHEET CONSTRUCTION

3.1.1. Standardised monthly bank balance sheet data[1] of South African banks are used from April 2015 to March 2017. The BA900 returns are not granular enough to allow the extraction of CET1 capital data, which was instead obtained from banks' annual statements, Pillar III capital disclosures and the Orbis Bank Focus database[2]. It is not sensible to use a period dating back further since the capital data becomes too scarce. Capital data before 2015 is difficult to obtain for all banks, since numerous small banks either did not exist, or their capital data for earlier periods is simply not available in the public domain.

3.1.2. As at March 2017 there were ten locally controlled banks, three mutual banks, six foreign controlled banks and fifteen branches of foreign banks, making up a total of 34 registered banks. For the purpose of this investigation, we do not consider the parent companies of the foreign branches. Firstly, subsidiaries may not be supported by the parent company. Secondly, while the assumption may under- or overestimate systemic risk in the local banking sector if subsidiaries were supported by the parent (depending on the solvency position of the parent company), it is necessary in order to keep the system closed. In other words, to ensure that risk levels within the system are not influenced by external market players, any actions that they may take or any regulations that may apply to them. This banking system can be considered as a typical candidate for a core-peripheral structure, as it consists of five large, 'core' banks and 29 smaller banks. To illustrate this, the total asset values of the banks are shown graphically in Appendix A, Figure A.1. For this study, eight banks are excluded because of a lack of capital data (more detail is given in ¶¶3.1.7 and 3.1.9 below), leaving 26 that are included in the analysis. The process for composing the simplified balance sheets as illustrated in Table 1 is explained below.

3.1.3 On the asset side, items are categorised according to whether they have a short, medium or long time to maturity at inception. Recall that short-term assets have a maturity of less than a month, medium-term assets have a maturity of more than a month and less than a year, and long-term assets have a maturity of more than a year. Assets that do not have a contractual maturity date are categorised according to their expected holding period, for example remittances in transit which are categorised as medium-term. Not all balance sheet items fall distinctly into only one category. Most of these items are placed into the category in which most individual assets are

---

expected to fall. For example, the local Treasury Bills can have maturities ranging from one day to twelve months, but normally have an unexpired maturity of 91 days or 182 days. Therefore, these are categorised as medium-term assets for our purpose. There are two exceptions to this rule:

– Marketable government stock on the BA900 forms (line item 198) is only given with a maturity of up to three years, and a maturity of over three years. Marketable government stock with a maturity of over three years are included in the long-term asset category. Marketable government bonds with a maturity of up to three years are assumed to be equally distributed across short-, medium- and long-term assets as all three of these maturity categories are included in this line item.

– Derivatives are divided according to term on the liability side of the BA900 forms, but not on the asset side. The assumption is therefore made that on the asset side of each bank, the proportion of short-term derivatives to total derivatives is the same as on the liability side. The same assumption holds for the medium- and long-term derivative instruments. If there are no derivatives on the liability side, the derivatives on the asset side are divided equally among the short-, medium- and long-term assets.

3.1.4   Derivative exposures constitute an important source of systemic risk because increased margining requirements during stress scenarios can place excessive strain on banks' liquidity positions. While this can be modelled using a network approach (see e.g. the study by Markose, Giansante & Shaghaghi (2012)), we do not explicitly model these exposures, but include such effects indirectly via the trust mechanism. This is because counterparty relationships are not publicly available, and it avoids complicating a model that is meant to remain simple.

3.1.5   Credit impairments with respect to loans and advances are deducted from the medium-term assets. This is because private sector loans and advances (that are categorised as medium-term assets) generally make up a large portion of total loans and advances and should also contain the majority of impaired accounts. Any impairments in respect of investments are deducted from the long-term assets since investments are generally regarded as long-term assets. The categorisation of assets is illustrated in Table B.1 in Appendix B.

3.1.6   Note that with more granular data the categorisation of assets according to maturity can be done more precisely. In this case it is necessary to aggregate the balance sheet items at this level since the available detail does not allow for a finer categorisation according to term. Regulators with more detailed information could use a larger number of categories so that assets can be grouped according to more time horizons and other characteristics as well.

3.1.7   A banks' Common Equity Tier I (CET1) capital represents the capital part of the balance sheet for the purpose of this investigation. This is because problems in financial systems can spread rapidly, and the CET1 capital can quickly be converted into cash (Gai & Kapadia 2010). (The same approach is used by Wells (2004), Mistrulli (2011) and Cont, Moussa & Santos (2012).) Additional Tier 1 capital is excluded since these must first be converted in the event of a crises. The equity side of the BA900 balance sheets is not sufficiently granular to allow for calculation of the banks' CET1 capital. For this reason, data from financial statements, published Pillar III capital disclosures and Orbis Bank Focus is used to supplement the primary balance sheet information. However, the data obtained via these sources are quarterly at best (in some cases only annually) and not all banks publish these on the same dates. Furthermore, some banks publish only risk weighted CET1 ratios and do not necessarily include a monetary amount for this type of capital. The available data for this part of the balance sheet must therefore be used to estimate the missing data points where possible.

3.1.8   To estimate a bank's monthly CET1 capital, the available CET1 amounts are divided by the corresponding total asset values of the respective bank at the available points in time. In other words, if $K_t$ and $A_t$ denote a bank's CET1 capital and total asset value at time $t$ respectively, we calculate $C_t = \frac{K_t}{A_t}$ for all months $t$. This gives an unweighted ratio of CET1 to total assets at selected points in time. There are two main reasons for using this ratio. It firstly strips out any inflationary effects over time and secondly removes the effect of significant increases or decreases in banks' growth rates. Where available, the unweighted ratio of CET1 to total assets are very stable for all banks over the period considered (the maximum variance for this ratio for over all banks is 0,00344). Therefore, for most banks the missing unweighted CET1 ratios could easily be estimated.

3.1.9   Table B.2 in Appendix B shows all the unweighted CET1 ratios for registered local banks that could be obtained from the available data. For each bank that has at least three CET1 data points available between May 2017 and February 2015, the remaining ratios are estimated for the outstanding months. Banks that have less than three CET1 data points are excluded from the analysis, reducing the total number of banks from 34 to 26. The total assets of all excluded banks make up less than 3% of all banks' assets as at May 2017.

3.1.10  For the remaining banks, the available unweighted CET1 capital ratios are used to estimate the unknown CET1 ratios as follows:

– Where missing data points fall in-between two known data points, linear interpolation between the two known data points is used to estimate the missing values. For example, if the ratios $C_\tau$ and $C_{\tau+3}$ are available for months $\tau$ and $\tau + 3$, but ratios for months $\tau + 1$ and $\tau + 2$ are not, we use the estimates $\hat{C}_{\tau+k} = C_\tau + \frac{k(C_{\tau+3} - C_\tau)}{3}$ for $k = 1, 2$.

– Where a missing data point does not lie between two known data points, the average unweighted CET1 ratio for the associated bank is taken. For example, if no CET1 data is available for month $t = 1$, then $\hat{C}_1 = \frac{1}{m} \sum_\tau C_\tau$, where $m$ is the number of months $\tau$ for which $C_\tau$ is available and the sum is taken over all available ratios $C_\tau$.

3.1.11  Once the estimates $\hat{C}_t$ for the CET1 capital are determined, all the required balance sheet entries are known. The next step is then to specify the interactions between the banks that are represented by the edges, where different assumptions regarding these interactions lead to different network structures.

## 3.2   NOTATION AND DEFAULT CASCADES

3.2.1   The modelling procedure is based on the work of May & Arinaminpathy (2010). In this section the month $t$ is fixed, and therefore subscripts relating to the month are not included as in section 3.1. Suppose a network consists of $N$ banks, where each bank $i$'s total assets are denoted by $a_i$. The short-, medium- and long-term assets of a bank $i$ are denoted by $a_i^{(s)}$, $a_i^{(m)}$ and $a_i^{(l)}$ respectively. Finally, bank $i$'s CET1 capital is denoted by $c_i$. For ease of reference, the terms CET1 capital and capital will be used interchangeably for the remainder of the paper.

3.2.2   We choose an initial bank $n$ and suppose that it suffers an initial loss. For the purpose of this paper, such an event is called an 'initial shock' since we assume that it was a significant and unexpected event. In this event bank $n$ loses a fraction, say $s$, of its assets. If $s \cdot a_n > c_n$, it fails, and the shortfall causes friction in the market. It is noted that the bank might not technically be insolvent at this point but might rather be in liquidation. However, for our purposes it is excluded from the network, and hence the distinction between liquidation and insolvency is not required.

3.2.3    Now three effects come into play. Firstly, the regulator may require other banks to assist with capitalisation in order to limit the spread of losses to other parts of the economy by making whole the retail and institutional creditors' unsecured loans. We assume that a proportion, say $u$, of this shortfall must be covered by the remaining banks. The remaining proportion $1 - u$ is absorbed by the Total Loss Absorbing Capacity (TLAC) part of the troubled bank's balance sheet, after which unsecured creditors bear the loss. The resulting funding requirement is spread over all banks in the system in proportion to their asset sizes. In other words, if bank $n$ experiences an initial loss event, its capital is reduced by $S_n = s \cdot a_n$. If $S_n \geq c_n$, then $n$ defaults and each bank $i$ suffers a loss of

$$L_i^{(1)} = u(S_n - c_n) \cdot \frac{a_i}{\sum a_k}.$$

3.2.4    Secondly, we include losses due to raised provisions and mark-to-market effects. For each remaining bank in the system, the reduced value of the short-term assets is given by

$$a_i^{(s)} \cdot \exp\left(-g^{(s)}\right),$$

where $g^{(s)}$ is a parameter associated with the reduction of value for the short-term assets. This is a commonly used method in the systemic risk literature for modelling changes in asset prices due to changes in supply and demand (Cifuentes, Ferrucci & Shin, 2005, May & Arinaminpathy, 2010, Gai & Kapadia, 2010, Nier et al., 2008).

3.2.5    The medium- and long-term assets are reduced in the same way, where the associated parameters are given by $g^{(m)}$ and $g^{(l)}$. This implicitly assumes that all banks in the system hold similar classes of assets, which is generally not the case. However, to avoid over-complicating the model we make the simplifying assumption that all banks will be affected to the same degree. Each of these parameters represents the expected effect that the insolvency of a bank would have on the assets in the system. It is referred to as liquidity losses or liquidity shocks for the remainder of this paper as the methodology is similar to the liquidity shocks presented by e.g. Nier et al., (2008). The associated parameters are referred to as liquidity reduction parameters. At this stage each bank $i$ in the system experiences a liquidity loss of

$$L_i^{(2)} = \sum_{\eta \in \{s, m, l\}} a_i^{(\eta)} \cdot \exp\left(-g^{(\eta)}\right).$$

3.2.6    Finally, we include losses due to a deterioration in market sentiment. The perceived exposure of other banks to the problems faced by bank $n$ determines the edges in the network. An edge starting at a bank $j$ and pointing towards bank $n$ means that the market believes $j$ may be exposed to similar difficulties as $n$, or may be adversely affected by the default of $n$. The edges in the network are assumed to be random, and the different structures are discussed further in section 3.3. Recall that the shortest distance between two nodes is the smallest number of edges that can be used to travel from the one to the other. The shortest distance $d(i, n)$ in the network from any bank $i$ to the failing bank $n$ determines the degree to which $i$ is affected by a loss of trust. Small values of $d(i, n)$ indicate 'closeness' in the network, which represents a perceived tendency for a bank $i$ to experience similar problems as $n$. In order to reflect the shrinkage of a bank $i$'s balance sheet due to increased funding costs and any resulting forced sales, each asset class of bank $i$ is reduced by a factor $\exp\left(-\frac{\delta}{d(i,n)}\right)$. Therefore, each remaining bank $i$ in the system experiences a further loss of

$$L_{i,n}^{(3)} = \sum_{\eta \in \{s,m,l\}} a_i^{(\eta)} \cdot \left[1 - \exp\left(-\frac{\delta}{d(i,n)}\right)\right], \tag{1}$$

where $\delta$ is the associated reduction factor. Similar to the parameters $g^{(\eta)}$ for $\eta \in \{s,m,l\}$, different proximity factors should be assigned to different types of assets when applying this model in practice. However, we avoid introducing too many parameters for the purpose of illustration by using the same parameter for all asset classes.

3.2.7    This type of loss is called a proximity shock for the remainder of the paper, as it is related to the distance between banks in the network and avoids confusion with losses arising from the devaluation of assets following the default of a bank. Proximity shocks aim to capture the consequences of the market's sentiment following a bank's default. The reaction of the market will typically depend on the circumstances surrounding the default and therefore it is preferred to take a generalised modelling approach to capture the effects of market sentiment. For example, following the curatorship of African Bank, some of the larger banks received credit downgrades from Moody's which increased their cost of borrowing. The reason given for the downgrade was that while the reserve bank did mitigate contagion risk by issuing a bailout, some creditors were allowed to suffer losses and hence Moody's was of the view that there is a "lower likelihood of systemic support from South African authorities to fully protect creditors in the event of need"[3]. Other examples include the banking crisis in Greece which saw a run on the banks that negatively affected banks' liquidity positions, and the European sovereign debt crisis which resulted in credit downgrades and increased costs of borrowing. Instead of restricting ourselves to particular scenarios, we consider a wide range of possibilities regarding the spread of distrust in the system. This is done by simulating network paths according to the structures presented in section 3.3, which avoids the need to consider the circumstances surrounding the initial default. This approach implicitly assumes that distrust in the system is only initiated after a default, and that banks of equal distance to the defaulting bank will experience losses of a similar degree. The appropriateness of the first assumption depends on the circumstances surrounding the default, and whether the market was aware of any friction within the system beforehand. The second assumption is unlikely to hold in practice but is required to keep the model simple and tractable. Finally, the model makes the underlying assumption that the loss of market sentiment due to a bank's failure is independent of the failed bank's size. In practice, it is expected that the failure of bigger banks will affect market sentiment more adversely than that of smaller banks. However, it is not straightforward to determine the magnitude of such differences, and the resulting effects may obscure the network implications that we aim to investigate in this paper.

3.2.8    The way that proximity shocks are modelled in the network accounts for the fact that some banks will experience a worse loss of confidence than others. From equation (1) it is seen that banks with smaller shortest distances to the failing bank will experience worse losses than those with greater shortest distances. This is illustrated in Figure 3 below, where the failing bank is indicated by the cross. The darker nodes experience greater losses than the lighter node, since they have a smaller shortest distance to the failing bank. The edges in the network is directed to take account of a wide range of possibilities without overcomplicating the model. In some circumstances the default of one bank may lead to distrust in another bank, but not the other way around. For example, the default of a large, systemically important bank may affect the market's perception of small banks, but the default of a small bank may not necessarily affect the perceived financial positions of much larger banks. Note that the direction of the edges does not represent the direction in which the losses spread but rather represents the similarity between banks. For example, if a directed edge exists from bank $i$ to bank $j$, the interpretation is that bank $i$ is similar to bank $j$ in the sense that the market perceives $i$ to be exposed to similar difficulties as bank $j$ in the event of bank $j$'s default.

---

[3] https://www.moodys.com/research/Moodys-downgrades-four-South-African-banks-on-review-for-further--PR_306571

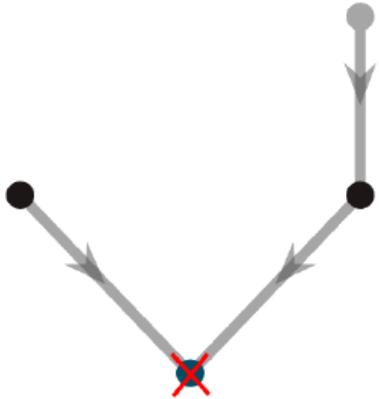

Figure 3. Illustration of a proximity shock following the default of a bank

3.2.9    The total loss to each remaining bank $i$ is given by $L_{i,n} = L_i^{(1)} + L_i^{(2)} + L_{i,n}^{(3)}$. The losses for each bank are subtracted from their respective capital amounts, leading to further bank failures whenever $L_{i,n} \geq c_i$. For each further bank failure, losses due to funding requirements, liquidity shocks and proximity shocks are calculated. For each remaining bank in the system, all losses are added together to determine the next round of failures. This is repeated until the default cascade stops, i.e. until either all banks have defaulted or the remaining banks in the system have absorbed all losses. Let $\theta_n$ denote the total number of banks that had defaulted because of the initial shock of bank $n$ (including bank $n$). The proportion of banks that have defaulted, say $\alpha_n = \frac{\theta_n}{N}$ is then calculated.

3.2.10  The above procedure is repeated for $n = 1,2, \ldots, N$. The average proportion of defaulted banks over the $N$ repetitions of the cascade is then calculated, i.e. we calculate $\alpha = \frac{1}{N}\sum_{n=1}^{N} \alpha_n = \frac{1}{N^2}\sum_{n=1}^{N} \theta_n$. This is repeated $m$ times, with the edges simulated each time. The average defaulted fraction over all the simulations is then denoted by $\bar{\alpha}$. For the remainder of the paper, we refer to $\bar{\alpha}$ as the systemic risk indicator.

## 3.3    NETWORK STRUCTURES

3.3.1    Recall that the network structure is determined by the way that banks are connected to one another in the network. We construct a network of trust deterioration, where the edges represent paths through which trust is lost in the system.

3.3.2    As it is not possible to know beforehand which banks will be perceived as being affected by another bank's failure, the edges in the network are assumed to be random and various network structures are considered. Even though some structures may be unrealistic, it is of interest to include them to consider a wider range of outcomes. This allows for a better understanding of the relevance of network structure in a network model based on trust deterioration.

3.3.3    Six network structures are investigated based on the probability $p(i, j)$ that a connecting edge from bank $i$ to bank $j$ is present. To take account of heterogeneity between banks, we let this probability be dependent on the relative asset sizes of banks. The structures described in this section are selected as they are either well known structures found in the network theory literature and capture a range of possibilities or facilitate the modelling of core-peripheral structures in networks that are not necessarily scale-free.

3.3.4    Figures 4 to 9 illustrate the behaviour of each structure. The size of a node is indicative of the bank's total asset value. A solid line represents a high probability that the associated edge is present, and a dashed, transparent line indicates a lower probability. The formulae for determining and standardising the connection probabilities $p(i, j)$ are included in APPENDIX C.

3.3.5    Figure 4 contains the first structure, which is an Erdös-Rényi network and is the simplest of all the structures considered. The probability that an edge exists between two nodes is independent of the asset sizes and is the same between all banks.

3.3.6    Figure 5 illustrates the second structure. Here, it is assumed that the probability that a large bank causes a loss of trust in any other bank is high. Shocks experienced by smaller banks have a small probability of affecting other banks. This structure is termed 'flight to quality'. Here, the market assumed before the shock event that the bigger banks were the most financially sound. The failure of a big bank therefore causes widespread panic, affecting most other banks in the system.

3.3.7    A disassortative network structure is included, where banks of dissimilar size are more likely to have connecting edges between them. This is illustrated in Figure 6. The assortative structure shown in Figure 7 exhibits the opposite behaviour, where banks of similar size are more likely to have connections between them. Such structures may not be realistic for banking networks. They are included to widen the range of structures and because of the prominent role that these networks play in related fields of study such as social networks or ecosystems where e.g. 'opposites attract'.

3.3.8    The final two structures represent core-peripheral networks, where a network has a small, highly connected core (the top tier), with a larger, sparsely connected peripheral (the bottom tier). As the South African system consists of a small number of big banks and several small banks, it is reasonable to include core-peripheral structures into our range of networks. The Tiered type I network is illustrated by Figure 8. Large banks have high probabilities of being linked to one another, and small banks lower probabilities of being connected to one another. The probabilities of large and small banks being connected to one another lie in between.

3.3.9    The final structure is termed Tiered type II and is more refined than the previous structures. As shown by Figure 9, the probability that:

–    a small bank connects to another small bank is low;
–    a small bank connects to a large bank is also relatively low;
–    a large bank connects to a small bank is high; and
–    a large bank connects to another large bank is also high.

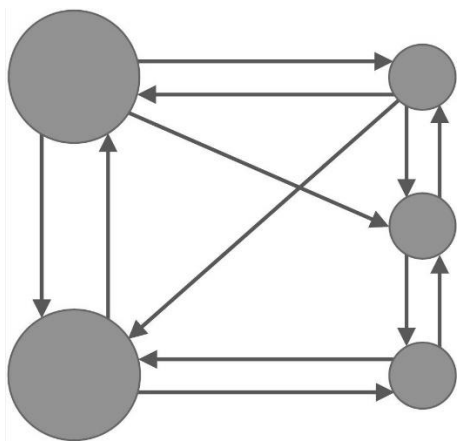

Figure 4. Illustration of an Erdös-Rényi structure's connection probabilities

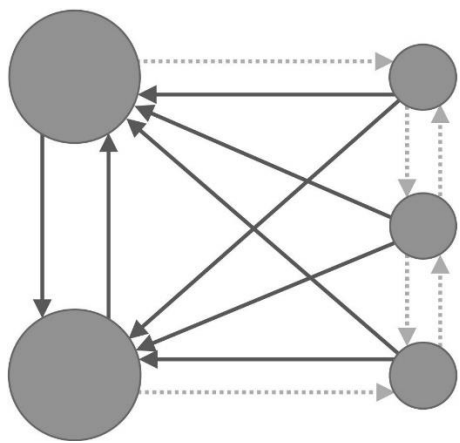

Figure 5. Illustration of a flight to quality structure's connection probabilities

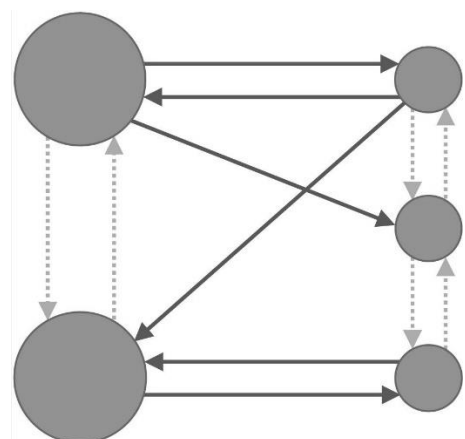

Figure 6. Illustration of a disassortative structure's connection probabilities

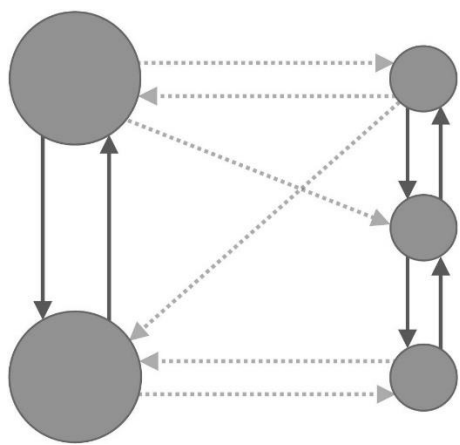

Figure 7. Illustration of an assortative structure's connection probabilities

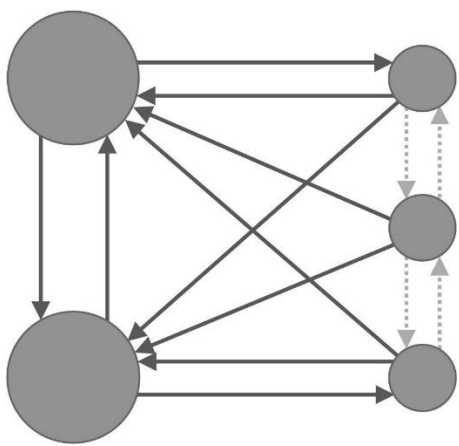

Figure 8. Illustration of a Tiered type I structure's connection probabilities

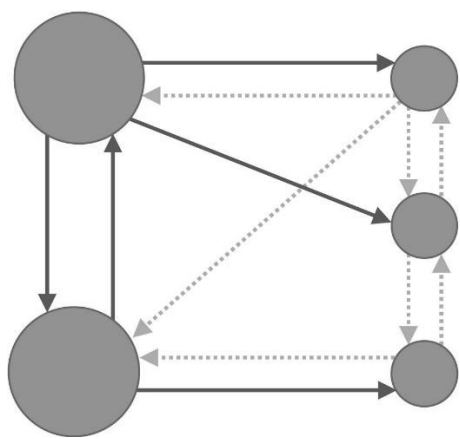

Figure 9. Illustration of a Tiered type II structure's connection probabilities

## 4. IMPLICATIONS FOR REAL WORLD BANKING SYSTEMS

### 4.1 APPLYING THE MODEL TO SOUTH AFRICAN BALANCE SHEET DATA

4.1.1.  The different network structures discussed in section 3.3 are compared to one another over time. The combined effect of network structure, the system's interconnectedness and the consequences of liquidity shortages and a deterioration of market sentiment on systemic risk is investigated. Recall that systemic risk is measured by calculating the probability that a bank defaults as a result of a shock to the system. For ease of reference, liquidity risk and the risk of loss due to a deterioration of market sentiment are referred to as indirect risk from here onwards. This is because losses resulting from these risks are not directly attributable to exposures between banks.

4.1.2.  We illustrate how the systemic default indicator changes over time by calculating a point in time probability at each month during the investigation period. At each time interval, an initial shock of 0,4 is applied to the system. In other words, the bank that suffers the initial loss as explained in section 3.2 experiences a loss equal to 40% of its total asset value. Whenever a bank defaults, it is assumed that 30% of the shortfall must be covered by the remaining banks, i.e. we assume that $u = 0,3$. Four scenarios are considered regarding the interconnectedness and the effect of indirect risk factors on systemic risk:

–  Low risk parameters $\left(g^{(s)} = 0,01, g^{(m)} = 0,01, g^{(l)} = 0,02 \text{ and } \delta = 0,01\right)$ and a moderate level of interconnectedness ($\bar{p} = 0,5$).
–  Low risk parameters $\left(g^{(s)} = 0,01, g^{(m)} = 0,01, g^{(l)} = 0,02, \text{ and } \delta = 0,01\right)$ with a high level of interconnectedness ($\bar{p} = 0,8$).
–  High values for the risk parameters $\left(g^{(s)} = 0,015, g^{(m)} = 0,015, g^{(l)} = 0,03, \text{ and } \delta = 0,015\right)$ with a moderate level of interconnectedness ($\bar{p} = 0,5$).
–  High risk parameters $\left(g^{(s)} = 0,015, g^{(m)} = 0,015, g^{(l)} = 0,03, \text{ and } \delta = 0,015\right)$ and a high level of interconnectedness ($\bar{p} = 0,8$).

4.1.3.  The values for $g^{(s)}, g^{(m)}, g^{(l)}$ and $\delta$ for the high and low indirect risk scenarios are simply a scaling of one another. Note that the parameter associated with the long-term assets is higher than for the other maturities. This is done to account for the illiquidity of these assets. It is important to note that different results may be obtained with different combinations of these parameters. However, it is impractical to consider an arbitrary number of combinations without more information regarding realistic values. Therefore only a few combinations are considered in this study.

4.1.4.  For the low-risk scenario, the chosen parameter values imply that after each default, the short- and medium-term assets of banks are decreased by approximately 1% and the long-term assets by 2%. The proximity shock parameter reduces all asset values of banks with a shortest distance of 1 to the failing bank by 1%. For the high-risk scenarios, the short- and medium-term assets are reduced by 1,5% and the long-term assets by 3%. The proximity shocks decrease all assets of banks directly connected to the failing bank by 1,5%.

4.1.5.  Figures 10 to 13 show the relative levels of systemic risk over time for all network structures considered in section 3.3. Where the graphs reach a flat baseline just below 0,04, the system did not experience any additional defaults over and above the initial default. In those cases, the average fraction of defaults experienced in the system is one out of 26. It is noted that the systemic default indicators shown by the graphs are based on hypothetical values of the parameters associated with indirect risks and are not necessarily accurate. This is because the focus of this study is on the relative risk levels associated with different network structures, and not to calculate actual probabilities for these events.

4.1.6.   As expected, higher levels of interconnectedness result in lower levels of discrimination between the different structures. The lines in Figures 11 and 13 are closer to one another when compared to Figures 10 and 12 respectively. This is because the higher value of $\bar{p}$ pushes the probabilities $p(i,j)$ to towards one for all structures and hence they become more representative of fully connected systems. The levels of risk over time does not change significantly when interconnectedness is increased, and the overall shapes of the graphs are preserved when increasing the level of connectedness of the system.

4.1.7.   From all four scenarios, it is seen that there is a spike in systemic risk around December 2015. This corresponds to the month during which former South African finance Minister Nhlanhla Nene was replaced, which was an unexpected and controversial political event in South Africa. The local financial market reacted negatively, and the local currency depreciated significantly during that period.

4.1.8.   A second spike in systemic risk is observed around June 2016. This increase is less prominent in Figures 12 and 13 (where higher risk parameters are used) than in Figures 10 and 11 (where lower risk parameters are used). This was also a time during which the Rand depreciated steeply against the US dollar. This was due a combination of factors, namely a weak economic growth outlook, rumours that the former finance Minister was to be arrested and an approaching credit review by Standard and Poor to decide whether they will downgrade South Africa's sovereign rating to junk status. During March 2017, former finance Minister Pravin Gordhan was also replaced during another controversial political event. This coincides with a sudden increase in systemic risk in Figures 10 to 13.

4.1.9.   The prominence of the December 2015 spike may be explained by looking at the average balance sheet items over time (see Figure A.2 in Appendix A). At December 2015, the relative increase in the average for the short- and long-term assets are much greater compared to the CET1 capital. This could, on average, lead to relatively larger losses for the initially shocked bank (because this is defined as a proportion of assets) that need to be absorbed by the capital. However, the average asset values at June 2016 and March 2017 do not show the same extreme behaviour, which could explain why these spikes are less prominent.

4.1.10.   However, Figures 12 and 13 with high risk parameters show a smaller increase in systemic risk. This may be because overall risk levels are higher in these scenarios, thereby decreasing the prominence of the spikes. This shows that the level of indirect risk influences which events lead to an increase in systemic risk.

4.1.11.   In general, it appears that the importance of the network structure is to a large extent influenced by the values chosen for the risk parameters. In the low indirect risk scenarios (Figures 10 and 11) at times when systemic risk levels are relatively high, the network structures exhibit small differences. Otherwise they are virtually indistinguishable from one another.

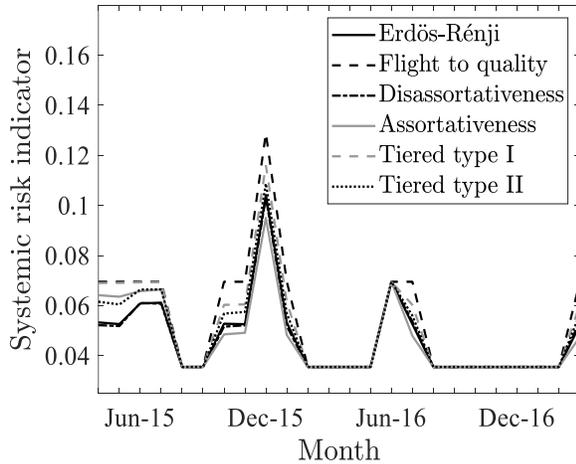

Figure 10. Low indirect risk, moderate interconnectedness

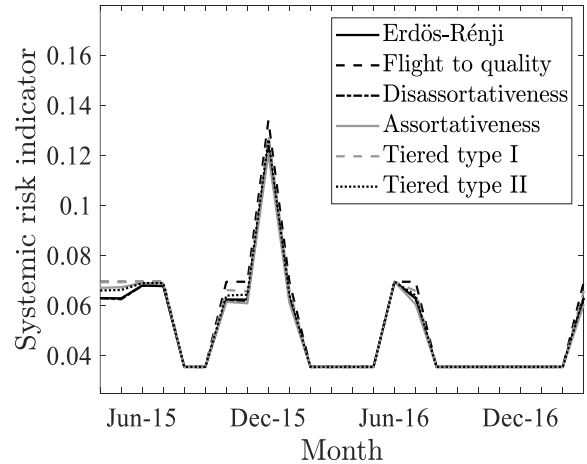

Figure 11. Low indirect risk, high interconnectedness

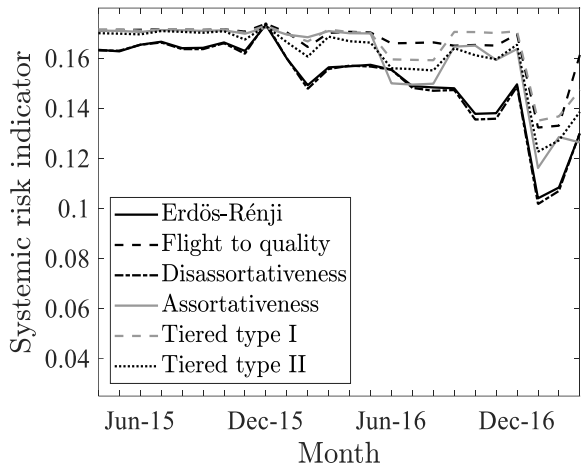

Figure 12. High indirect risk, moderate interconnectedness

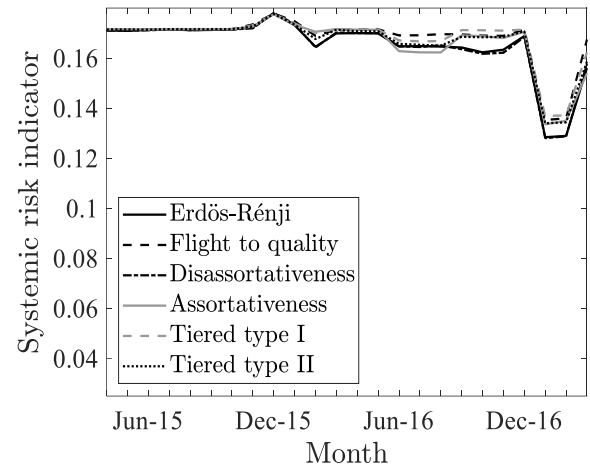

Figure 13. High indirect risk, high interconnectedness

4.1.12. For higher risk parameters (Figures 12 and 13), the network structure plays a greater role in the level of systemic risk. Overall, the flight to quality structure exhibits the most risk when differences between the structures can be seen. From all four scenarios it is seen that the structures are mostly consistent regarding directional changes, i.e. the structures' risk levels move in the same direction at each time step, albeit at different rates. The only exception is around September 2016 in Figures 12 and 13, where the Erdös-Rényi and disassortativeness structures show a slight decrease in risk, whereas the other structures show an increase.

4.1.13. The above results show that the indirect risk parameters can affect how systemic risk changes over time. To further illustrate this point, we consider the effect of changing the relative values of the indirect risk parameters. A base parameter value of 0,015 is used for all indirect risk parameters. These parameter values will be referred to as the base parameters for the remainder of the section. The resultant graph of systemic risk over time is shown in Figure 14.

4.1.14. The effect of increasing any one risk parameter is considered. The liquidity risk parameters are each increased to 0,03, where the proximity shock parameter is increased to 0,025.

This is because the results are very sensitive to this parameter, which is reasonable as it affects assets of all maturities. Therefore, increasing it to 0,03 increases the risk levels too much. The level of interconnectedness is kept at 0,5, since it was seen above that increasing the interconnectedness does not significantly influence the shape of the graphs. Instead, it brings the structures closer to one another.

4.1.15. Figure 15 shows the effect of increasing only the parameter associated with the short-term liquidity losses from 0,015 to 0,03. Figures 16 to 18 show the same results for the medium-term, long-term and proximity shock parameters, respectively. Note that Figure 17 is the same as Figure 12 but is included again and scaled to facilitate the comparison between graphs.

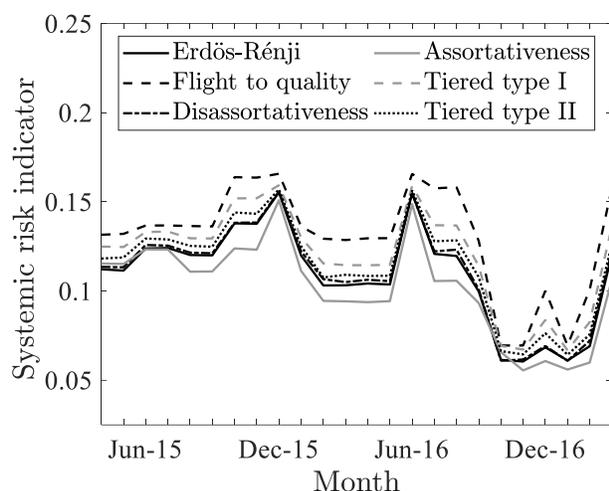

Figure 14. Systemic risk over time for base parameters of $g^{(s)} = 0{,}015$, $g^{(m)} = 0{,}015$, $g^{(l)} = 0{,}015$ and $\delta = 0{,}015$

4.1.16. From Figures 15 to 18 it is seen that the parameters don't have the same effect on systemic risk. By increasing only the short-term liquidity parameter in Figure 15, the peaks in systemic risk are more pronounced than in Figure 14 for the base parameters. Differences between the network structures are decreased at the peaks but are more pronounced at the troughs. Increasing only the medium-term liquidity parameter (Figure 16) flattens out the graph to such an extent that the December 2015 and June 2016 spikes are not distinguishable from other peaks in the graph. Only the December 2016 decrease in risk is preserved. Once again, the differences between the network structures become greater during the December 2016 dip in risk but are less during other months when compared to Figure 14.

4.1.17. When only the long-term liquidity risk parameter is increased in Figure 17, the graph again flattens out to an extent, but the dip in risk levels during December 2016 is preserved. Differences between the network structures are generally more pronounced than in Figures 15 and 17.

4.1.18. By increasing only the parameter associated with market sentiment, the general level of risk increases quicker than for the other parameters. The December 2015 peak becomes much more pronounced than in Figures 15 to 17. Despite this parameter being directly related to the network structure, the differences between the structures become less. This suggests that the effect of the increased emphasis on the parameter related to the network structure is diminished by the increase in defaults experienced by all structures.

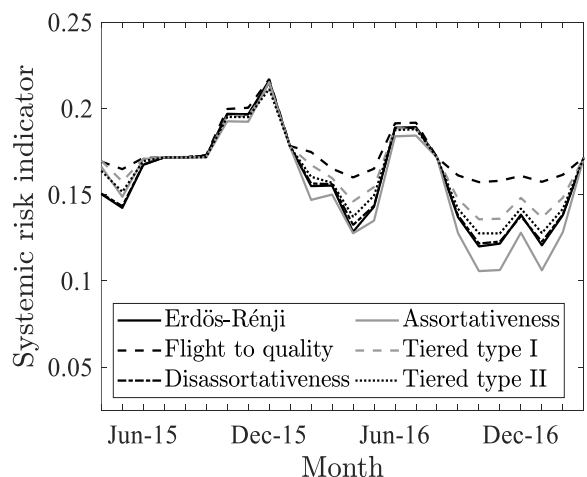

Figure 4. Systemic risk over time for base parameters, but with $g^{(s)} = 0{,}03$

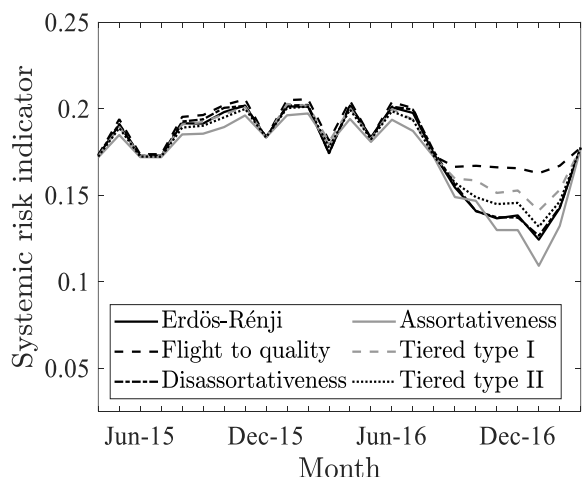

Figure 5. Systemic risk over time for base parameters, but with $g^{(m)} = 0{,}03$

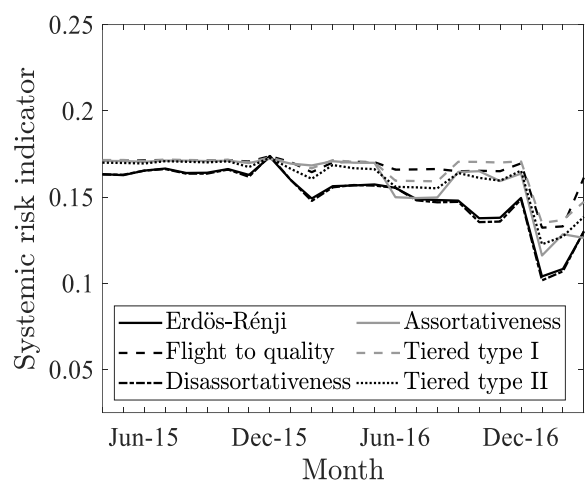

Figure 6. Systemic risk over time for base parameters, but with $g^{(l)} = 0{,}03$

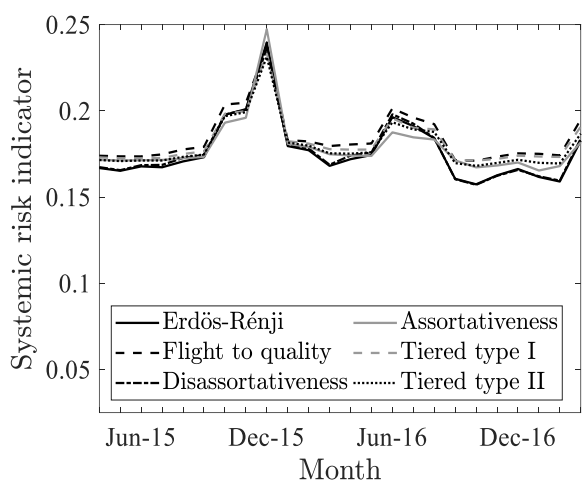

Figure 7. Systemic risk over time for base parameters, but with $\delta = 0{,}025$

4.1.19. The differences between Figures 15, 16 and 17 are likely because of differences in asset values for different maturities between banks and within each bank. This is because the three liquidity parameters enter the model in the same way via reductions in the associated asset values. Therefore, networks derived from different countries' banking systems will likely differ in the way that they react to changes in network structure and liquidity risk parameters. For regulators, it is important to note that conclusions reached for one banking system will not necessarily hold for another.

4.1.20. The results show that both network structure and indirect risk are important in determining the level of risk present in the system. Network structure can affect the degree to which market disturbances fuel systemic risk. Determining parameters associated with liquidity risk for different asset types and market sentiment is important for network models of systemic risk, since these can significantly influence results.

4.1.21. To understand how the network structure affects how the different banks in the system contribute to systemic risk, it is useful to consider how $\alpha_n$ (defined in ¶3.2.10 as the proportion of nodes that default if $n$ is the initially shocked bank) varies with the asset value of bank $n$. Let $\bar{\alpha}_n$ denote the average of $\alpha_n$ over 2000 simulations. It is reasonable to expect the default of larger banks to have a greater knock-on effect on the system compared to smaller banks and hence $\bar{\alpha}_n$ is expected to be higher for larger banks. This is confirmed by Figures 19 and 20 below, which are based on the Erdös-Rényi network structure, with Figure 20 based on balance sheet data as at March 2017. Figure 19 shows the average value of $\bar{\alpha}_n$ for large, medium, small and very small banks. Figure A.1 in Appendix A was used to determine the groups. The four largest banks in the system are included in the 'large' group, the fifth largest bank in the 'medium' group, the sixth to thirteenth largest banks (Capitec Bank to African Bank) in the 'small' group, and the remainder in the 'very small' group. Figure 20 shows a scatterplot of $\bar{\alpha}_n$ against the logarithm of $n$'s asset value. Both figures support the expectation that large banks have a greater knock-on effect when they default. It is interesting to note that all the other structures lead to the same conclusions (the graphs are omitted to avoid repetitiveness). This shows that the model tends to behave as expected in this regard irrespective of the network structure.

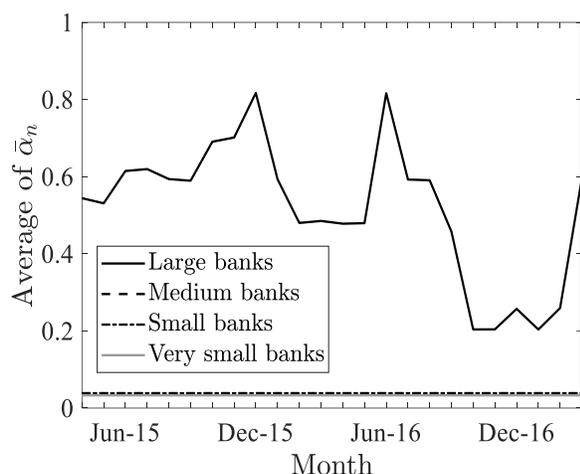

Figure 19. Systemic risk indicator by bank size for the Erdös-Rényi network

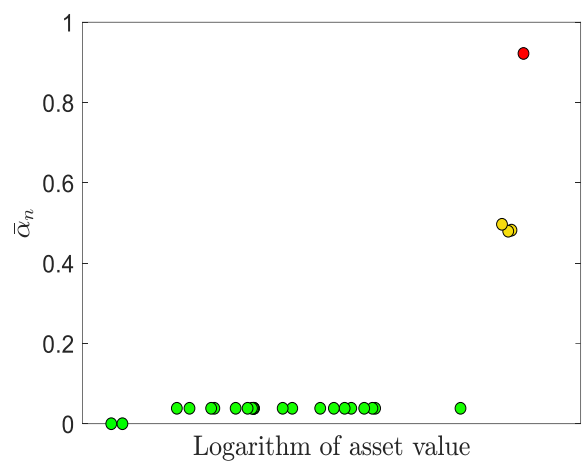

Figure 20. Scatterplot of $\bar{\alpha}_n$ against the natural logarithm of bank $n$'s assets for the Erdös-Rényi network

## 4.2 IMPLICATIONS OF RESULTS

4.2.1. The network structures behaved similarly for most of the cases considered here. The levels of risk were generally similar for high levels of connectivity and for lower levels of indirect risk. However, the differences in risk levels between the structures were not consistent over time. There were many time periods when the risk levels were very close to one another. Where the structures' risk levels did differ from one another, they mostly followed similar trends over time. This suggests that the changes in systemic risk detected by the model is not highly dependent on the network structure. The general relationship between a bank's size and its contribution to systemic risk was also the same for the different structures. These observations have the following implications:

(1) The materiality of network structure is firstly influenced by the objective of the network model. If the objective is to accurately determine the level of risk in the system, then the network structure does not make a significant difference for highly interconnected systems. For lower levels of interconnectedness, there are time periods when the network structures exhibit similar risk levels. However, this is highly dependent on the risk parameter values.

(2) If the objective is to detect changes in systemic risk, the materiality of network structure decreases. This means that the uncertainty around the initial shock to the system (and hence the resulting path through which losses spread) is less problematic.

(3) Network models are capable of capturing the intuitive relationship between a bank's size and the consequences of its default for a wide range of network structures.

4.2.2. The results show that the risk parameters significantly influence the extent to which network structure affects systemic risk. The liquidity risk mechanism employed by the model affects the assets of all banks and therefore is not directly related to the network structure. Nevertheless, small changes in the liquidity risk parameters have a non-trivial influence on the relative differences in risk exhibited by the structures.

4.2.3. Furthermore, each risk parameter influences the results in its own way. For example, increasing the short-term liquidity parameter emphasised the December 2015 and June 2016 increases in risk, whereas the other liquidity parameters reduced the significance of these spikes. This either means that the model is not able to detect increases in risk for certain parameter values, or that the system does not experience a significant increase in systemic risk during times of market turmoil for some liquidity scenarios. For example, the high liquidity risk scenarios (see Figures 12 and 13) might increase the risk levels during all months to such an extent that the effect of weak economic conditions are diminished.

4.2.4. The above observations have the following implications for the modelling of systemic risk using a network approach:

(1) Empirical studies that aim to determine the level of systemic risk should take care to calibrate the liquidity risk parameters to levels appropriate for the system being considered. The difficulty of calibrating these parameters is a drawback of such a network approach to modelling systemic risk. As incidents of bank closure/liquidation can be scarce, one may have to work with few datapoints. As such it would not be possible to calibrate the parameters precisely, but it could be possible to determine a realistic range for the required parameters. One can consider the balance sheet positions of all banks before and after each closure/liquidation incident to measure the size of any shrinkages in the remaining banks' balance sheets.

(2) A finer division of assets is recommended. The fact that the liquidity risk parameters each had a different effect on the model output suggests that the classification of assets can be a material aspect of such a study.

(3) Since the model showed increases in systemic risk during times of market turmoil, it shows that network models of systemic risk may be valuable modelling tools. A great advantage of this is that publicly available balance sheet information can be used to model systemic risk, thereby avoiding the need to obtain confidential trading information. It may be by chance that the model detected increases in systemic risk due to balance sheet fluctuations.

This warrants further investigation to determine with greater certainty whether the model can accurately identify potential crises.

## 5.    CONCLUSION

5.1    We use a novel network approach to model systemic risk in South Africa by considering how the liquidity problems and default of one bank can lead to the market frictions such as losing trust in the financial wellbeing of other banks. Here, the type of event that leads to the default of the first bank is likely to infer the network structure. As this cannot be determined beforehand, and a lack of past systemic crises make liquidity parameters difficult to determine, we consider the effect of network structure and liquidity risk on the results.

5.2    The network structure's influence on systemic risk was considered under different circumstances and over time. The general trend is the same for all network structures, showing that the model may detect fluctuations in systemic risk even if the true network structure is unknown. The differences between the network structures are influenced by the effect of liquidity risk and the losses due to negative investor sentiment. The trends in systemic risk over time is sensitive to changes in the parameters associated with these risks. This shows that any investigation regarding systemic risk in banking networks must incorporate indirect losses such as losses due to liquidity problems and a deterioration of market sentiment, since these significantly influence the results.

5.3    The effect of indirect losses has a significant effect on how the system reacts to changes in structure and interconnectedness. This indicates that the calibration of these parameters is important when making decisions based on network models of systemic risk. The importance of this is emphasized by the fact that systemic risk levels over time behave significantly different depending on the combination of all the parameters used. It is imperative or regulators to incorporate and accurately model these effects when assessing the effect of proposed regulatory changes.

5.4    Despite the problems associated with determining the correct network structure and liquidity risk parameters, such models can be useful. These models are simple, easy to understand and makes use of publicly available balance sheet data. The framework presented here can be useful to answer 'what-if' questions that arise in practice and to give insight into what might happen to the system given an appropriate network. The framework in itself can be used to generate a wide range of output, for example one can investigate a range of different risk measures (average capital lost, average proportion of asset value lost by the system etc., and test for correlations between this and the size of the initially defaulted bank.

5.5    During the time frame considered here, the network model detected increases in systemic risk at times when the economy experienced unexpected market disturbances. An important avenue for future research is to determine whether this is by chance, or whether the model accurately predicts the probability that a crisis can occur. Furthermore, it is important to note that the model does not forecast times of distress, but instead provides a proxy for the level of risk at a point in time. In other words, the true proportion of banks defaulting following a shock to the system is not determined, but rather a value that increases or decreases along with it.

5.6    Note that this investigation solely focused on relative changes in systemic risk and is not meant to provide an accurate estimate of the probability that a bank defaults following an initial shock to the system. Important avenues for future research include assigning banks different probabilities of receiving an initial shock and including the possibility that such a shock may affect more than one bank at the same time. This should be done parallel to modelling the macro-economic environment over time. In this case a lender of last resort and the cost of regulation can be included. Furthermore, the effect of leverage can be modelled more explicitly in the presence of a macro-economic environment. Finally, the interaction between the banking sector and other financial institutions (e.g. insurance and investment companies) should be included in future work. For further work, it is important to consider a range of different risk measures and to test whether the same conclusions hold. Other important directions for future work include considering a wider range of structures, a finer division of assets and using a networks-on-networks approach to incorporate a range of contagion mechanisms.

## ACKNOWLEDGEMENTS

The financial assistance of the National Research Foundation (NRF) and the Absa chair in Actuarial Science towards this research is hereby is hereby acknowledged. Opinions expressed, and conclusions arrived at, are those of the authors and are not necessarily to be attributed to the NRF or Absa. The authors further acknowledge the valuable advice and input given by Jeannette de Beer.

APPENDIX A.

ADDITIONAL FIGURES

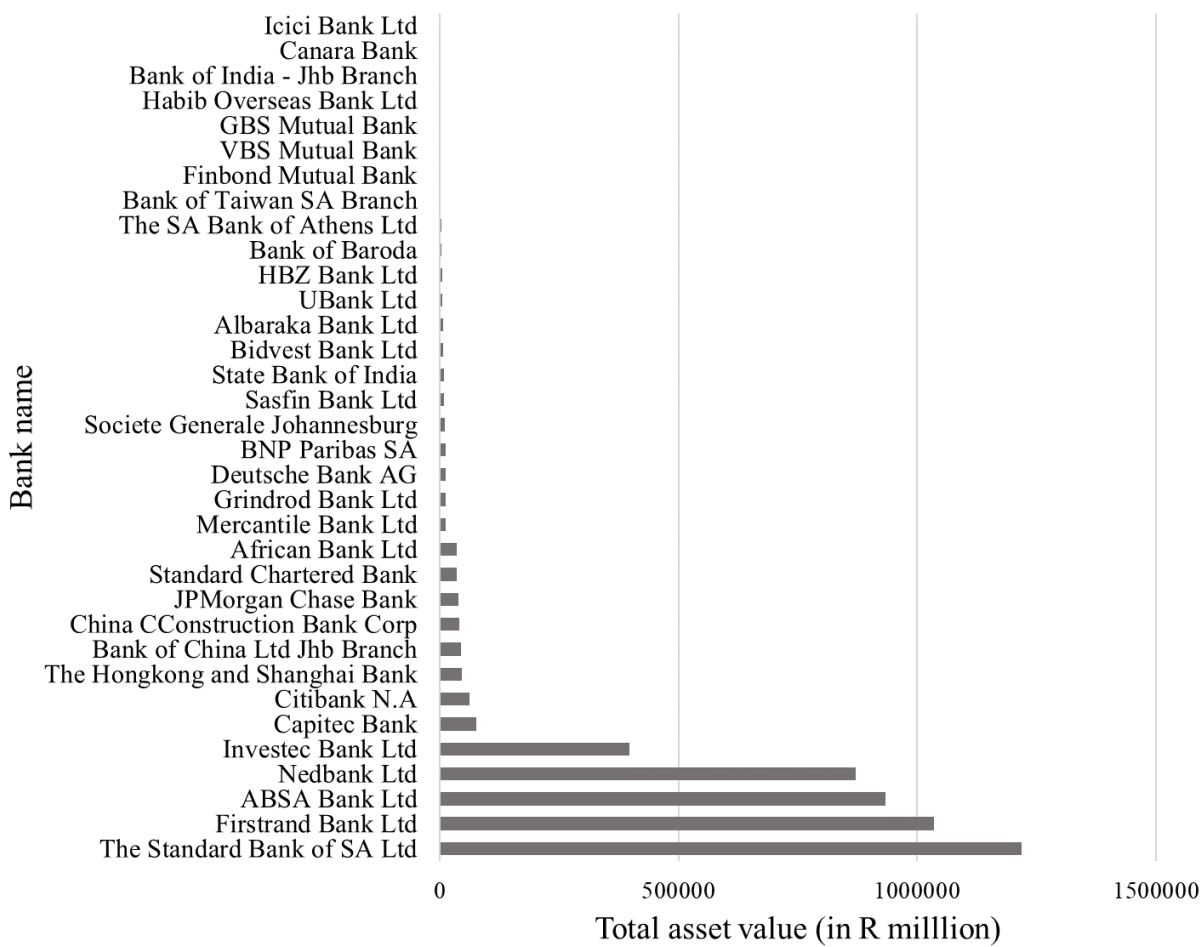

Figure A.1. The distribution of assets in the banking sector as at 31 March 2017

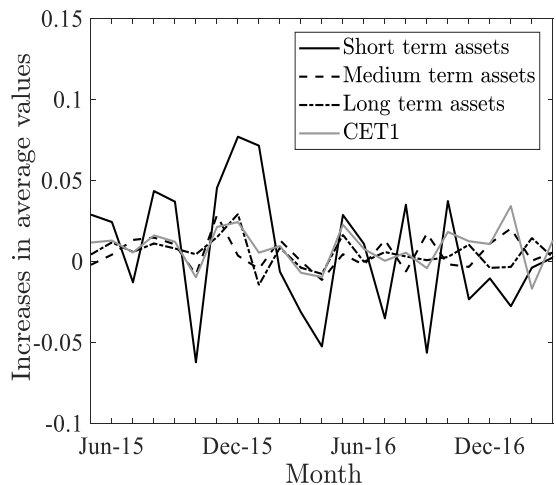

Figure A.2. Relative increases of the average balance sheet items in the system

APPENDIX B.

BALANCE SHEET INFORMATION

Table B.1. Division of assets according to term

**Short Term Assets**

– Central bank money and gold.
– Deposits with, and loans and advances to banks.
– Loans granted to the SARB and other institutions under resale agreements.
– Foreign currency deposits, loans and advances.
– One third of marketable government stock that have an unexpired maturity of less than 3 years.
– Derivative instruments assigned to the short-term asset category according to the rules in section 3.1.

**Medium Term Assets**

– Instalment sales.
– Credit-card debtors.
– Overdrafts, loans and advances to the private sector.
– Bankers' acceptances (Treasury bills, SARB bills, promissory notes, commercial paper and Land Bank bills).
– Clients' liabilities per contra.
– Remittances in transit.
– Current income tax receivables and deferred income tax assets.
– One third of marketable government stock that have an unexpired maturity of less than 3 years.
– Derivative instruments assigned to the medium-term asset category according to the rules in section 3.1.

**Long Term Assets**

– Redeemable preference shares.
– Leasing transactions.
– Mortgage advances.
– Overdrafts, loans and advances to the public sector.
– Non-marketable government stock.
– All marketable government stock excluding two thirds of those stock that have an unexpired maturity of less than 3 years.
– Debentures and other interest-bearing security investments of private sector.
– All equity investments.
– Derivative instruments assigned to the long-term asset category according to the rules in section 3.1.
– Other investments.
– Non-financial assets.
– Retirement benefit assets.
– Assets acquired or bought to protect an advance or investment.

Table B.2. Available unweighted ratios of CET1 to total assets at month-end for all registered banks

| Bank Name | 03/2017 | 02/2017 | 12/2016 | 09/2016 | 06/2016 | 03/2016 | 02/2016 | 12/2015 | 09/2015 | 06/2015 | 03/2015 | 02/2015 |
|---|---|---|---|---|---|---|---|---|---|---|---|---|
| | | | | | | | | | | Reporting month | | |
| ABSA Bank Ltd | 0.0642 | | 0.0656 | 0.0561 | 0.0540 | 0.0525 | | 0.0538 | 0.0525 | 0.0489 | 0.0495 | |
| African Bank Ltd | 0.2387 | | 0.2259 | 0.2178 | | | | | | | | |
| Albaraka Bank Ltd | | | 0.1057 | 0.1070 | 0.1094 | 0.1111 | | 0.1083 | | 0.1075 | 0.1089 | |
| Bank of Baroda | 0.0459 | | 0.0578 | | 0.0410 | | | | | | | |
| Bank of China LTD Jhb Branch | | | | | | | | | | | | |
| Bank of India - Jhb Branch | | | 0.6626 | 0.7247 | 0.6865 | 0.6471 | | 0.5647 | 0.6692 | 0.6947 | 0.6733 | |
| Bank of Taiwan SA Branch | | | | | | | | | | | | |
| Bidvest Bank Ltd | 0.2483 | | 0.2439 | 0.2544 | 0.1745 | 0.1895 | | 0.1929 | 0.2051 | 0.1664 | 0.2009 | |
| BNP Paribas SA | | | | | | | | 0.0439 | | | | |
| Canara Bank | 0.8010 | | 0.7844 | 0.7367 | 0.6515 | 0.6810 | | 0.6468 | 0.6796 | 0.7160 | 0.7774 | |
| Capitec Bank | | 0.1990 | | | | | 0.1989 | | | | 0.1956 | |
| China Construction Bank | | | 0.1063 | 0.0838 | 0.0549 | 0.0618 | | 0.0407 | 0.0485 | 0.0482 | | |
| Citibank N.A | 0.0945 | | 0.0887 | | 0.0981 | 0.0960 | | 0.0568 | 0.0701 | 0.0794 | 0.0926 | |
| Deutsche Bank AG | 0.1087 | | 0.1204 | 0.1007 | 0.0907 | 0.0836 | | 0.0603 | 0.0658 | 0.0719 | 0.0589 | |
| Finbond Mutual Bank | | 0.2085 | | | | | 0.2407 | | | | 0.2425 | |
| Firstrand Bank Ltd | 0.0704 | | 0.0672 | 0.0693 | 0.0688 | 0.0663 | | 0.0674 | 0.0688 | 0.0681 | 0.0642 | |
| GBS Mutual Bank | | | | | | | | | | | | |
| Grindrod Bank Ltd | 0.0665 | | 0.0506 | 0.0654 | 0.0640 | 0.0702 | | 0.0571 | 0.0749 | 0.0725 | 0.0719 | |
| Habib Overseas Bank Ltd | 0.1146 | | 0.0994 | 0.0965 | 0.0917 | 0.0862 | | 0.0826 | 0.0779 | 0.0787 | 0.0840 | |
| HBZ Bank Ltd | 0.0749 | | 0.0785 | 0.0865 | 0.0741 | 0.0710 | | 0.0662 | 0.0646 | 0.0696 | | |
| Icici Bank Ltd | | | | | | | | | | | | |
| Investec Bank Ltd | 0.0854 | | 0.0857 | 0.0827 | 0.0813 | | | 0.0836 | 0.0836 | 0.0878 | | |
| JPMorgan Chase Bank | 0.1313 | | 0.1136 | 0.0840 | 0.0721 | 0.0656 | | 0.0568 | 0.0849 | 0.0961 | 0.0785 | |
| Mercantile Bank Ltd | 0.1714 | | 0.1649 | 0.1729 | 0.1789 | 0.1837 | | 0.1903 | 0.1964 | 0.1971 | 0.2039 | |
| Nedbank Ltd | 0.0594 | | 0.0575 | 0.0544 | 0.0556 | 0.0545 | | 0.0547 | 0.0524 | 0.0542 | 0.0572 | |
| Sasfin Bank Ltd | 0.1491 | | 0.1671 | 0.1836 | 0.1976 | 0.2070 | | 0.2090 | 0.1975 | 0.1975 | 0.1944 | |
| Societe Generale Jhb | | | | | | | | | | | | |
| Standard Chartered Bank | | | | 0.0911 | 0.0874 | 0.0829 | | 0.0789 | 0.0957 | 0.0972 | 0.0830 | |
| State Bank of India | 0.1722 | | | | | 0.1252 | | 0.1265 | 0.1323 | 0.1357 | 0.1394 | |
| The Hongkong and Shanghai Bank | | | | | | | | | | | | |
| The SA Bank of Athens Ltd | 0.0851 | | 0.0844 | 0.0889 | 0.0767 | 0.0803 | | 0.0802 | 0.0928 | 0.1005 | 0.1053 | |
| The Standard Bank of SA Ltd | 0.0570 | | 0.0552 | 0.0556 | 0.0570 | 0.0546 | | 0.0550 | 0.0536 | 0.0549 | 0.0518 | |
| Ubank Ltd | | | | | | | | | | | | |
| VBS Mutual Bank | | 0.0857 | | 0.0878 | 0.0878 | 0.0878 | 0.0832 | | | | 0.0973 | 0.0877 |

APPENDIX C.

CONNECTION PROBABILITIES ASSOCIATED WITH DIFFERENT NETWORK STRUCTURES

C.1.    Let $p(i,j)$ be the probability that bank $i$ has an outgoing edge to bank $j$ and let $a_j$ denote the asset value of bank $j$. As banks cannot be connected to themselves, $p(i,j) = 0$ whenever $i = j$. For the rest of this discussion, assume that $i, j \in \{1, 2, \ldots, N\}$, with $i \neq j$. The different network structures are then specified as follows:

1.  Erdös-Rénji:

$$p(i,j) = p$$

2.  Flight to quality:

$$p(i,j) = \frac{a_j}{\max_k \{a_k\}}$$

3.  Disassortativeness:

$$p(i,j) = \frac{\max \left\{ \frac{a_i}{a_j}, \frac{a_j}{a_i} \right\}}{\max_{k \neq m} \left\{ \frac{a_k}{a_m}, \frac{a_m}{a_k} \right\}}$$

4.  Assortativeness:

$$p(i,j) = \frac{\min\{a_i, a_j\}}{\max\{a_i, a_j\}}$$

5.  Tiered type I:

$$p(i,j) = \frac{a_i + a_j}{\max_{k \neq m} \{a_k + a_m\}}$$

6.  Tiered type II:

$$p(i,j) = \frac{a_i + a_j + \max\{a_i - a_j, 0\}}{3 \cdot \max\{a_k\}}$$

C.2.    For a given set of asset values, the average number of edges will differ between the different structures. For comparative purposes, it is important to work with similar levels of connectedness between the different structures. In addition to this, it is important to have a systematic way of varying the level of connectedness in the network to investigate its effect on systemic risk.

C.3.    For these reasons, a method for scaling the probabilities is required. Such a method will need to provide scaled probabilities that remain in the range $[0,1]$. It must further allow one to standardise the different structures to represent the same level of connectivity between them. The level of connectivity is measured by means of the average probability that an edge exists in the system (i.e. the average probability that one bank is exposed to a loss of trust in case of default of another).

C.4.     Let $\bar{p}_0$ be the average probability that an edge exists in the system based on any one of the above six structures. Then

$$\bar{p}_0 = \frac{\sum_{i=1}^{N} \sum_{\substack{j=1 \\ j \neq i}}^{N} p(i,j)}{N(N-1)}.$$

Note that the entries for which $i = j$ are all zero are not included in the calculation of this average, and they are to remain zero after scaling. Let $\bar{p}$ be the desired average connection probability. The $p(i,j)$ probabilities then needs to be scaled in order to yield this average. Suppose $\bar{p}_0 > \bar{p}$. Then we choose

$$\bar{p}(i,j) = p(i,j)\frac{\bar{p}}{\bar{p}_0}$$

and if $\bar{p}_0 < \bar{p}$, then

$$\bar{p}(i,j) = 1 - \big(1 - p(i,j)\big)\frac{1-\bar{p}}{1-\bar{p}_0}.$$

The new connection probabilities $\bar{p}(i,j)$ will then have the required average $\bar{p}$.